\documentclass[journal]{IEEEtran}

\usepackage{graphicx}
\usepackage[cmex10]{amsmath}
\usepackage{amsfonts}
\usepackage{amssymb}
\usepackage{cite}

\begin{document}


\title{Superconducting-semiconductor quantum devices: from qubits to particle detectors}

\author{Yun-Pil Shim and Charles Tahan%
\thanks{Yun-Pil Shim is with the Department of Physics, University of Maryland, College Park, MD, 20742 USA, %
and also with Laboratory for Physical Sciences, College Park, Maryland 20740, USA %
e-mail: (ypshim@lps.umd.edu).}%
\thanks{Charles Tahan is with Laboratory for Physical Sciences, College Park, Maryland 20740, USA %
e-mail: (charlie@tahan.com)}%
\thanks{Manuscript received June xx, 2014; revised xx xx, 2014.}}


\maketitle

\begin{abstract}
Recent improvements in materials growth and fabrication techniques may finally allow for superconducting semiconductors to realize their potential. Here we build on a recent proposal to construct superconducting devices such as wires, Josephson junctions, and qubits inside and out-of single crystal silicon or germanium. Using atomistic fabrication techniques such as STM hydrogen lithography, heavily-doped superconducting regions within a single crystal could be constructed. We describe the characteristic parameters of basic superconducting elements---a 1D wire and a tunneling Josephson junction---and estimate the values for boron-doped silicon. The epitaxial, single-crystal nature of these devices, along with the extreme flexibility in device design down to the single-atom scale, may enable lower-noise or new types of devices and physics. We consider applications for such super-silicon devices, showing that the state-of-the-art transmon qubit and the sought-after phase-slip qubit can both be realized. The latter qubit leverages the natural high kinetic inductance of these materials. Building on this, we explore how kinetic inductance based particle detectors (e.g., photon or phonon) could be realized with potential application in astronomy or nanomechanics. We discuss super-semi devices (such as in silicon, germanium, or diamond) which would not require atomistic fabrication approaches and could be realized today.
\end{abstract}


\IEEEpeerreviewmaketitle


\section{Introduction}

The emergence of superconductivity in heavily-doped semiconductors has long been predicted \cite{cohen_rmp1964}. In recent years, group-IV semiconductors such as diamond \cite{ekimov_nature2004}, silicon \cite{bustarret_nature2006}, and germanium \cite{herrmannsdorfer_prl2009} have been successfully doped with acceptors reaching high enough hole density to turn superconducting. The more recent experiments in silicon are of significant interest as they show highly uniform, highly predictable superconductivity (versus density), while still maintaining the expitaxial nature of the crystal \cite{blase_nmat2009}. With respect to device physics, a recent proposal has raised the possibility of using novel material preparation and atomistic fabrication techniques \cite{fuhrer_nanolett2009,fuechsle_nnano2010} to realize superconducting (SC) devices in purified silicon crystals \cite{shim_tahan_ncomm2014}.

Superconducting circuits are very versatile for various classical and quantum device applications, from sensors to computation. The Josephson junction offers a loss-less non-linearity of particular relevance to quantum information processing and technology \cite{SC_qubit_review}. Qubits based on advanced circuit designs such as the transmon \cite{blais_pra2004} have reached remarkably long coherence times \cite{transmon,paik_prl2011} and continue to improve. To realize a practical quantum computer, still better gate fidelities are needed to implement quantum error correction toward fault-tolerant quantum computing. In present-day, heterogeneous superconducting devices (metals, insulators, substrates), device performance is limited by loss, often at the surfaces or interfaces \cite{oh_prb2006,pappas1,pappas2,sendelbach_prl2009}. Although it is unclear what the ultimate limits of loss in superconducting silicon might be, that devices could be constructed inside a single, purified crystal may offer improved performance someday.

Compared to conventional superconductor devices, super-silicon devices could be fabricated with atomic precision (though that is not required). Recently developed scanning tunneling microscope (STM) lithography \cite{fuhrer_nanolett2009,fuechsle_nnano2010} has been used to create electronic devices down to the single dopant level \cite{fuechsle_nnano2012}. This STM lithography technique (combined with atomic layer doping) has been used only for electron doping so far, reaching very high density of electrons, up to one in four crystal atoms replaced with a dopant \cite{weber_science2012}. There is high motivation to extend this technique for acceptor doping \cite{rusko_prb2013,shim_tahan_ncomm2014}.  

In this paper, we theoretically investigate basic characteristics of super-silicon for device applications and propose a few possible applications. We consider two types of qubits, the now ``standard'' transmon~\cite{transmon} and the more elusive phase-slip flux qubit \cite{mooij_harmans_njp2005}, which may be a natural fit for this system. Also of potential interest for device applications, we calculate the kinetic inductance (KI) of super-silicon, show that it is naturally high, and consider its use for KI particle detectors, of potential interest to astronomy (photon, phonon) and nanomechanics applications. We note that there are device geometries that won't require precision doping, which may enable other materials---most intriguingly superconducting diamond or whole wafers of super-silicon---to be utilized for experiments today.

\section{Superconductivity Characteristics}

In this section, we refine the estimation of SC characteristic parameters from \cite{shim_tahan_ncomm2014}. We start with only experimental data and assumptions that the holes can be described by free electron like energy dispersion and that it is a Bardeen-Cooper-Schrieffer (BCS) \cite{BCS} superconductor in dirty local limit. The effective mass of hole bands can be obtained from numerical simulations or from experiments, but the usually cited values are for low density cases. In super-silicon, the hole density is much higher ($\gtrsim 10^{21}$ cm$^{-3}$) and the effective mass significantly overestimates the Fermi energy and it leads to some inconsistencies in estimating SC parameters. So we don't use the low density effective mass and derive the effective mass from experimental data regarding superconductivity. The highest observed critical temperature for boron-doped silicon (Si:B) is $T_\text{c}$=0.6K for hole density $n_\text{h}$=$4\times 10^{21}$cm$^{-3}$ \cite{marcenat_prb2010}. The upper critical field at zero temperature $B_\text{c2}(0)$=0.1T and the normal resistivity was measured to be $\rho_\text{n}$=100$\mu\Omega$cm. 

Since we are mostly interested in low temperature quantum applications ($T \ll T_\text{c}$), zero temperature parameters will be calculated. The Ginzburg-Landau (GL) coherence length is obtained from $B_\text{c2}(T)$=$\Phi_0/2\pi\xi^2_\text{GL}(T)$ where $\Phi_0$=$h/2e$ is the SC flux quantum,
\begin{equation}
\xi_\text{GL}(0) = \sqrt{\frac{\Phi_0}{2\pi B_\text{c2}(0)}} = 57\text{nm} .
\end{equation}
The effective penetration depth $\lambda$ can be calculated from $\mu_0 \lambda^2(0)$=$\hbar \rho_\text{n} / (\pi \Delta_0)$ where $\Delta_0$ is the SC energy gap at zero temperature, which is estimated to be $\Delta_0$=1.76$k_\text{B} T_\text{c}$=91$\mu$eV, 
\begin{equation}
\lambda(0) = \sqrt{\frac{\hbar\rho_\text{n}}{\mu_0\pi\Delta_0}} = 1.35 \mu\text{m}. 
\end{equation}
The GL parameter $\kappa_\text{GL}$=$\lambda(0)/\xi_\text{GL}(0)$=23.6, which is consistent with the observed type II superconductivity in super-silicon. The mean free path $l_\text{m}$ is obtained from an expression that can be derived from $l_\text{m}$=$v_\text{F} \tau$ and $1/\rho_\text{n}$=$n_\text{h}e^2\tau/m_\text{h}$ where $v_\text{F}$ is the Fermi velocity of holes, $\tau$ is the relaxation time, and $m_\text{h}$ is the hole effective mass;
\begin{equation}
l_\text{m}=(3\pi^2)^{1/3} \frac{\hbar}{e^2 \rho_\text{n} n_\text{h}^{2/3}} = 5.04\text{nm}.
\end{equation}
The BCS coherence length $\xi_0$ and London penetration depth $\lambda_\text{L}(0)$ are obtained from $\xi_\text{GL}(0)/\xi_0$=$\pi \lambda_\text{L}(0) / 2\sqrt{3}\lambda(0)$ and $\lambda(0)$=$\lambda_\text{L}(0) (\xi_0/l_\text{m})^{1/2}$,
\begin{eqnarray}
\xi_0 &=& \frac{12}{\pi^2} \frac{\xi^2_\text{GL}(0)}{l_\text{m}} = 793\text{nm},\\
\lambda_\text{L}(0) &=& \lambda(0) \left(\frac{l_\text{m}}{\xi_0}\right)^{1/2} = 108\text{nm}.
\end{eqnarray}
The effective mass $m_\text{h}$ is now calculated from $\xi_0$=$\hbar v_\text{F}/\pi\Delta_0$ and $v_\text{F}$=$\hbar k_\text{F}/m_\text{h}$=$\hbar (3\pi^2 n_\text{h})^{1/3} /m_\text{h}$;
\begin{equation}  
m_\text{h} = \frac{\hbar^2 (3\pi^2 n_\text{h})^{1/3}}{\pi\Delta_0\xi_0} = 1.65m_\text{e}.
\end{equation}
This effective mass is about three times larger than the low density heavy hole band effective mass. It gives a Fermi energy of $\varepsilon_\text{F}$=0.557eV, which is more consistent with the Fermi energy from simulations for a high hole density \cite{bourgeois_apl2007}. The thermodynamic critical field is 
\begin{equation}
B_\text{c}(0) = \frac{B_\text{c2}(0)}{\sqrt{2}\kappa_\text{GL}} = 3\text{mT}.
\end{equation}
The estimated mean free path $l_\text{m}$ is much less than $\xi_\text{GL}(0)$ and $\lambda(0)$, which is consistent with the assumption of super-silicon being in the local limit.

\section{Tunneling Josephson Junction}

\begin{figure}[!t]
  \centering
  \includegraphics[width=\linewidth]{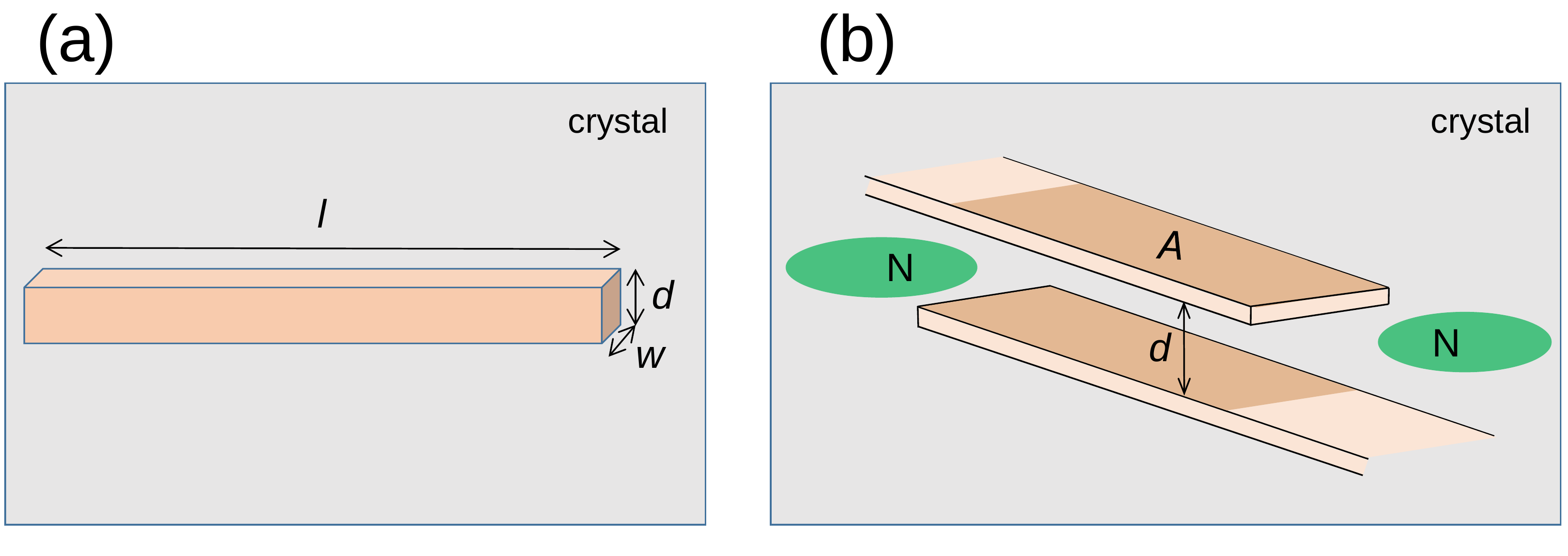}\\
  \caption{Basic superconducting elements for applications. The orange regions represent the extent of the impurity wave functions inside the crystal which will go superconducting when a sufficient density is reached. 
           (a) Superconducting wire. (b) Tunneling Josephson junction with additional gates. 
           They are both inside a single semiconductor crystal.}
  \label{fig:SC_wire_JJ}
\end{figure}

Now we define a few characteristic parameters for the tunneling JJ and estimate their values for the Si:B device.
All estimations will be for the hole density $n_h$=4.0$\times 10^{21}$cm$^{-3}$ and the SC parameters determined in the previous section. A JJ is typically characterized by a capacitance $C_J$ and an inductance $L_J$. The capacitance is simply the geometric capacitance given by 
\begin{equation}
C_\text{J} = \varepsilon_\text{r} \varepsilon_0 \frac{A}{d}.
\end{equation}
where the dielectric constant $\varepsilon_\text{r} \simeq 12$ for silicon. 
The JJ inductance $L_\text{J}$ is given by 
\begin{equation}
L_\text{J} = \frac{\hbar}{2e I_\text{c}}  = \frac{\Phi_0}{2\pi I_\text{c}},
\end{equation}
where $I_\text{c}$ is the JJ critical current given by \cite{ambegaokar_prl1963},
\begin{equation}
I_\text{c} = \frac{\pi \Delta(T)}{2e R_\text{n}} \tanh \left[ \frac{\Delta(T)}{2k_\text{B} T} \right].
\end{equation}
$R_\text{n}$ is the normal tunneling resistance, which is the inverse of the tunneling conductance $G_\text{n}$.
$G_\text{n}$ is proportional to the area $A$ and can be obtained by
\begin{equation}\label{eq:conductance}
\frac{G_\text{n}}{A} = \frac{m_{\text{h}}e^2}{2\pi^2\hbar^3} \int_0^{\varepsilon_{\text{F}}} d\varepsilon_{\text{z}} T(\varepsilon_{\text{z}}) 
\end{equation} 
where $T(\varepsilon_{\text{z}})$ is the transmission coefficient of a square barrier of height $V_\text{b}$ and width $d$,  
\begin{equation}
T(\varepsilon_{\mathrm{z}}) = \frac{4 \varepsilon_{\mathrm{z}} \left( V_{\mathrm{b}} - \varepsilon_{\mathrm{z}} \right)}{ 4 \varepsilon_{\mathrm{z}} \left( V_{\mathrm{b}} - \varepsilon_{\mathrm{z}} \right) + V_{\mathrm{b}}^2 \sinh^2 \kappa d },
\end{equation}
and $\kappa$=$\sqrt{\left( V_{\mathrm{b}} - \varepsilon_{\mathrm{z}} \right) 2 m_{\mathrm{h}}/\hbar^2}$.

\begin{figure}[!t]
  \centering
  \includegraphics[width=\linewidth]{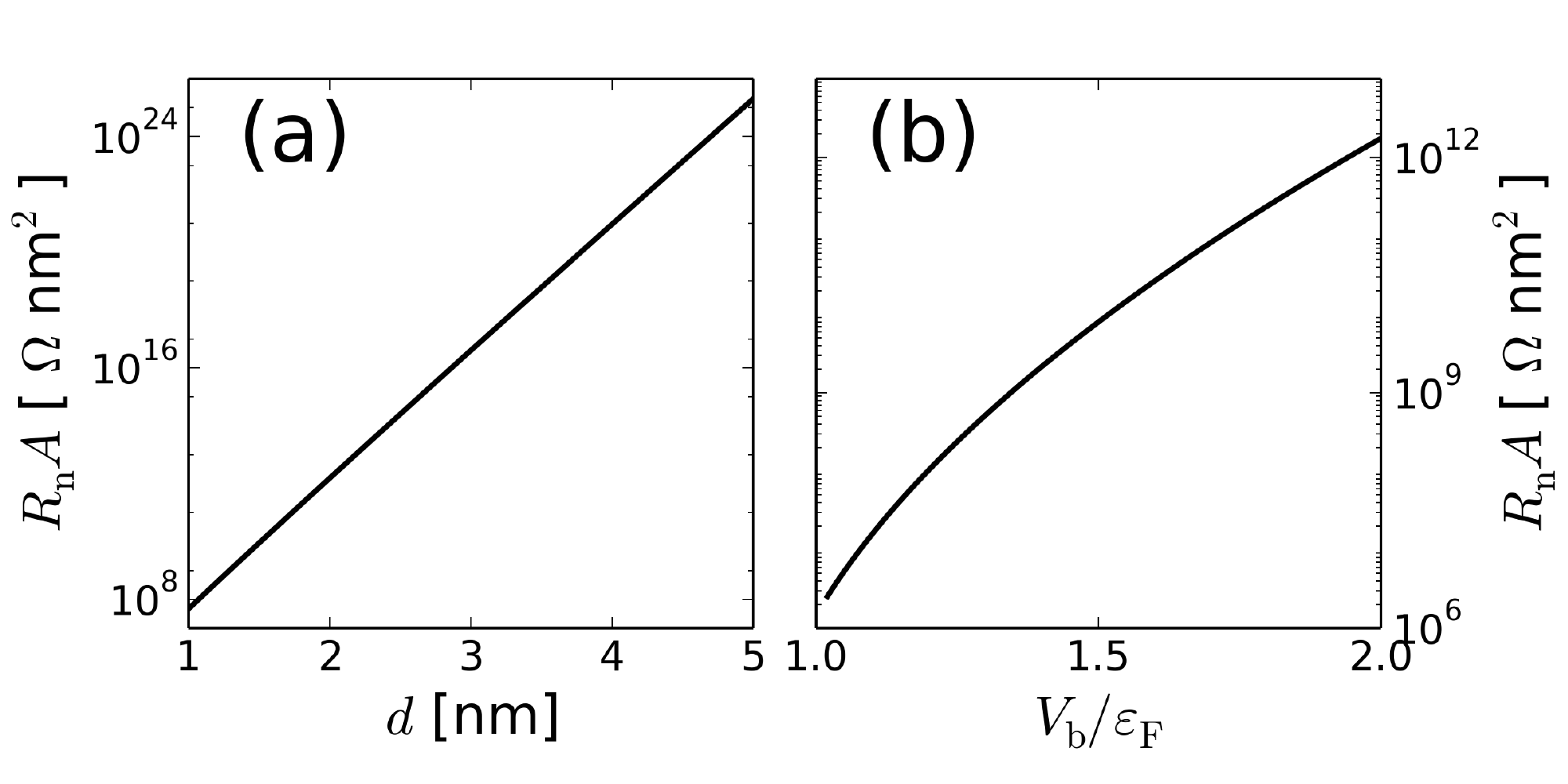}\\
  \caption{Tunneling resistance $R_\text{n}$ as a function of $d$ and $V_\text{b}$ across the silicon Josephson tunnel junction. 
           (a) For varying distance $d$ for $V_\text{b}$=$\varepsilon_\text{F}+E_\text{g}/2 \simeq$ 1.988$\varepsilon_\text{F}$. 
           $m_\text{h}$=1.65$m_\text{e}$. 
           (b) For varying barrier height $V_\text{b}$ with fixed $d$=2nm.}
  \label{fig:Tunneling_resistance}
\end{figure}

A crude estimate for the barrier height of the undoped region is $V_\text{b}$=$\varepsilon_\text{F}+E_\text{g}/2$. Figure \ref{fig:Tunneling_resistance}(a) shows the tunneling resistance as a function of the distance $d$, for $m_\text{h}$=1.65$m_\text{e}$ and $V_\text{b}$=$\varepsilon_\text{F}+E_\text{g}/2$. The tunneling resistance increases exponentially with increasing distance $d$ and the barrier height $V_\text{b}$. The barrier distance $d$ is controlled in the crystal growth process. On the other hand, the barrier height $V_\text{b}$ can be tuned by, e.g., additional gates such as doped regions with hole density less than the critical density for the SC transition (normal metal) as was depicted in Fig.~\ref{fig:SC_wire_JJ}(b), connected to outside voltages. The two characteristic energies of the JJ are
\begin{eqnarray}
E_\text{C} &=& \frac{(2e)^2}{2 C_\text{J}} = 3.02 [\text{eV} \cdot \text{nm}] \frac{d}{A} ~, \\
E_\text{J} &=& \frac{\hbar I_\text{c}}{2e} = \frac{\pi\hbar}{4e^2} \frac{\Delta(T)}{R_\text{n}} \tanh \left[ \frac{\Delta(T)}{2k_\text{B} T} \right] \\
    &\approx &  2.94 \times 10^{-1} [\text{eV} \cdot \mathrm{\Omega}] \frac{1}{R_\text{n}} ~,
\end{eqnarray}
where the last line is for zero temperature.

Another important parameter is the JJ plasma frequency which is defined as
\begin{eqnarray}\label{eq:JJ_omegap}
\omega_\text{p} &=& \frac{\sqrt{2 E_\text{C} E_\text{J}}}{\hbar} = \sqrt{\frac{2e}{\hbar}\frac{d}{\varepsilon_r\varepsilon_0} J_\text{c}}  \nonumber\\
 &=& 2.02\times 10^{15} [\text{Hz}\cdot\mathrm{\Omega}^{1/2} \cdot\text{nm}^{1/2}] \sqrt{\frac{d}{R_\text{n} A} }~,
\end{eqnarray}
where $J_\text{c}$=$I_\text{c}/A$ is the critical current density.    

As an example for the tunneling JJ, we take $A$=1$\mu$m$^2$, $d$=2nm, $V_\text{b}$=1.988$\varepsilon_\text{F}$. Then
$C_\text{J}$=53.13 fF, $L_\text{J}$=3.54 $\mu$H, $J_\text{c}$=93.1 A/m$^2$, $E_\text{C}$=6.03 $\mu$eV, $E_\text{J}$=0.191 $\mu$eV, and $\omega_\text{p}$=2.31GHz. By changing the geometry ($A$ and $d$) and the barrier $V_\text{b}$, we can tune these parameters to suit various applications.

\section{Kinetic Inductance of a Superconducting Wire}

Superconducting wires can be used in many parts of a circuit. They can be a lossless connecting wire between different elements, an inductor, a resonator, or even a phase slip junction. One of the potentially very useful properties of the SC semiconductors is their high kinetic inductance. Consider a SC wire with length $l$, width $w$, and depth $d$, as depicted in Fig.\ref{fig:SC_wire_JJ}(a). The kinetic energy due to the supercurrent, from GL description, is
\begin{equation}
E_\text{K} = \frac{1}{2} m^* v_\text{s}^2 n_\text{s}^* d w l \equiv \frac{1}{2} L_\text{K} I_\text{s}^2,
\end{equation}
where the supercurrent $I_\text{s}$=$n_\text{s}^* (2e) v_\text{s} w d$. Here $m^*$, $n_\text{s}^*$, and $v_\text{s}$ are effective mass, density, and velocity of a Cooper pair, from GL picture. The kinetic inductance $L_\text{K}$ is \cite{annunziata_nanotech2010}
\begin{equation}\label{eq:LK_GL}
L_\text{K} = \frac{m^*}{4 n_\text{s}^* e^2} \frac{l}{wd} = \mu_0 \lambda^2 \frac{l}{wd},
\end{equation}
where $\lambda$=$\sqrt{m^*/(4\mu_0e^2n_\text{s}^*)}$ is the effective penetration depth. An alternative way of finding an expression for the kinetic inductance based on BCS \cite{BCS} and Mattis-Bardeen \cite{Mattis_Bardeen} theories is from the complex conductivity $\sigma=\sigma_1-i\sigma_2$ of the superconductor, by equating $\sigma_2 (wd/l)$ with $1 / \omega L_\text{K}$. In the low frequency limit,  
\begin{equation}\label{eq:LK}
L_\text{K} = \frac{\hbar}{\pi} \frac{l}{wd} \frac{\rho_\text{n}}{\Delta(T) \tanh \left(\frac{\Delta(T)}{2k_\text{B} T} \right)}.
\end{equation}
This expression is valid for all range of temperature. At zero temperature, 
\begin{equation}\label{eq:LK0}
L_\text{K} (T=0) = \frac{\hbar}{\pi} \frac{l}{wd} \frac{\rho_\text{n}}{\Delta_0}  \simeq 0.18 \frac{\hbar R_\text{n}}{k_\text{B} T_\text{c}}, 
\end{equation}
and near the critical temperature $T \lesssim T_\text{c}$, it reduces to the GL result, (\ref{eq:LK_GL}). We characterize the low temperature kinetic inductance of various SC materials, by comparing the material parameter $\hbar\rho_\text{n}/\pi\Delta_0$ (see Table \ref{tab:kinetic_inductance}). The actual kinetic inductance will be determined by the geometry of the wire. 
The critical temperature $T_\text{c}$ and the resistivity $\rho_\text{n}$ are from the references cited and $\Delta_0$ was calculated using BCS relation $\Delta_0$=1.76$k_\text{B} T_\text{c}$. Kinetic inductance $L_\text{K}$ is proportional to $\hbar\rho_\text{n}/\pi\Delta_0$, from (\ref{eq:LK0}), and we calculated this geometry-independent material parameter for different SC materials.
Super-silicon has a kinetic inductance comparable to the most promising SC materials for applications requiring high kinetic inductance. Combined with the potential for much narrower quantum wire through STM lithgraphy, super-silicon could be very useful for phase slip junctions and kinetic inductance detector devices. We will give a few examples in the following sections. In Table \ref{tab:kinetic_inductance}, we also estimate the kinetic inductance for hole-doped superconducting diamond and germanium, which are also of promise for various device applications. 

\begin{table}[!t]
\caption{Kinetic inductance for various superconducting materials. 
}
\label{tab:kinetic_inductance}
\begin{tabular}{ccccc}
\hline
material & $T_\text{c}$ [K]  & $\Delta_0$ [$\mu$eV] & $\rho_\text{n}$ [$\mu \Omega \cdot$ cm]  & $\hbar \rho_\text{n}/ \pi\Delta_0$ [nH nm] \\
\hline\hline
Nb \cite{McCambridge_dissert} & 9.2 & 1395 & 10 & 0.015 \\
NbTi \cite{McCambridge_dissert} & 8.5 & 1289 & 74 &  0.12 \\
NbN \cite{gao_hajenius_apl2007} & 11.8 & 1790 & 240 & 0.28 \\
NbSi \cite{calvo_daddabbo_jltp2014} & 1.05 & 159 & 500 & 6.58 \\
TiN \cite{leluc_bumble_apl2010} & 4.1 & 622 & 100 & 0.34 \\
NbTiN \cite{barends_hortensius_apl2008} & 14.8 & 224 & 1700 & 1.59 \\
\hline
C:B \cite{ishizaka_prl2007} & 7.0 & 1062 & 680 &  1.34 \\
Si:B \cite{marcenat_prb2010} & 0.6 & 91 & 100 & 2.30 \\
Ge:Ga \cite{skrotzki_ltp2011} & 0.43 & 65 & 100 & 3.21 \\
\hline
\end{tabular}
\end{table}

\section{Applications for Quantum Computation}

\subsection{Transmon Qubit}

\begin{figure}[!t]
  \centering
  \includegraphics[width=\linewidth]{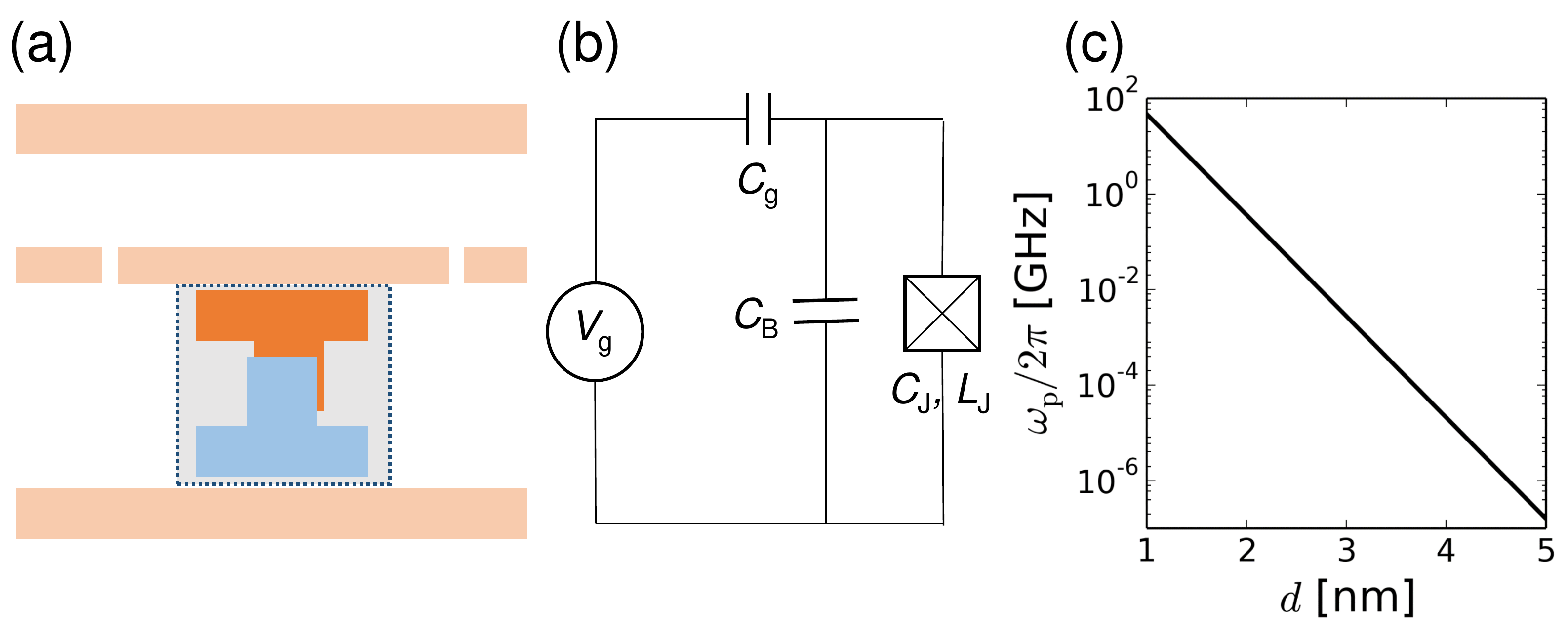}\\
  \caption{Transmon charge qubit. 
           (a) A schematic of a transmon charge qubit, coupled to a transmission line coplanar waveguide. 
               The JJ inside the the dashed square is the transmon qubit. In the super-semiconductor approach considered here, the transmon is formed from embedded superconducting regions inside the silicon (or germanium) and the tunnel junction (gap between the superconducting islands) is pure silicon crystal. The readout transmission line or cavity may be formed from super-silicon or superconducting metal on the surface.
           (b) Equivalent circuit of the transmon qubit.
           (c) The JJ plasma frequency $\omega_\text{p}$ as a function of the tunneling barrier distance $d$ for $V_\text{b}/\varepsilon_\text{F}$=1.988.}
  \label{fig:transmon}
\end{figure}

The transmon qubit \cite{transmon,transmon_exp} is probably the most promising qubit implementation using SC circuits so far.  It is a type of Cooper pair box (CPB) charge qubit designed to be insensitive to the charge noise which is the main cause of decoherence in the charge qubit. The Hamiltonian is 
\begin{equation}
H_\text{transmon} = E_\text{C} \left( \hat{n} - n_\text{g} \right)^2 - E_\text{J} \cos \hat{\phi},
\end{equation}
where $\hat{n}$ and $\hat{\phi}$ are the Cooper pair number operator and the phase of the SC order parameter, respectively. $n_\text{g}$ is the effective offset charge that can be tuned by applying external voltage $V_\text{g}$ to the coupling capacitor $C_\text{g}$ [see Fig.\ref{fig:transmon}(b)]. By tuning the parameters such that $E_\text{J} \gg E_\text{C}$ by increasing the total capacitance with a large shunting capacitor, the qubit levels become essentially flat as functions of the effective gate charge $n_\text{g}$ making the qubit states insensitive to the charge fluctuations in $n_\text{g}$. One consequence of making $E_\text{J}$ much larger than $E_\text{C}$ is that the anharmonicity of the energy levels is reduced. So there is a compromise and typically $E_\text{J} \simeq$ 10 is chosen, where the anharmonicity is still about 10\%, large enough to allow qubit operations. This is accomplished by increasing the capacitance with large superconducting metals attached to the JJ. The qubit states energy difference approaches the JJ plasma frequency $\hbar\omega_\text{p}$=$\sqrt{2 E_\text{C} E_\text{J}}$. The qubit is coupled to a resonator and the strong coupling regime, where the qubit-resonator coupling is larger than the decay rates, could be achieved in this circuit QED architecture taking advantage of the small mode volume of the quasi-1D transmission line resonator and/or the large dipole matrix elements between qubit states \cite{wallraff_nature2004,paik_prl2011}. This strong coupling with resonating modes is utilized for gate operations and measurements. 

Now we estimate a geometry suitable for a transmon. If we don't want to use a shunting capacitor for the JJ as is typical, and instead use a ``parallel-plate'' design as is possible with our epitaxial approach, then for a given target frequency $\omega_\text{p}$, we can get required $d/R_\text{n} A$ from (\ref{eq:JJ_omegap}). Then from the relation between $R_\text{n} A$ and $d$ [see Fig.~\ref{fig:Tunneling_resistance}(a)], we can read the necessary $d$ and $R_\text{n} A$. Since $\omega_\text{p}$ does not depend on the junction area $A$, we can directly find the necessary junction barrier distance $d$, as in Fig.~\ref{fig:transmon}(c). In addition to the target frequency, we want to tune $E_\text{J}$ and $E_\text{C}$ to be in the transmon regime. This determines optimal junction area. For example, for $\omega_\text{p}/2\pi$=5GHz and $E_\text{J}/E_\text{C} \simeq$ 10, the necessary geometry for the JJ is $d \simeq$ 1.46nm and $A$=0.95$\mu$m$^2$, so a large junction area is needed which comes from the requirement of large capacitance. If we use a shunting capacitance $C_\text{B}$ [see Fig.~\ref{fig:transmon}(b)], from target frequency $\omega_\text{p}$ and $E_\text{J}/E_\text{C}$ we obtain necessary values for $E_\text{C}$ and $E_\text{J}$, which can be tuned separately using the shunting capacitance. For $\omega_\text{p}/2\pi$=5GHz and $E_\text{J}/E_\text{C} \simeq$ 10, we need $E_\text{J}$=$10E_\text{C}$=46.2$\mu$eV. To allow a smaller $A$, we need smaller $d$. If we choose $d$=1.2nm, we need $A$=6.07$\times$10$^4$nm$^2$. Then $C_\text{J} \simeq$ 5.38aF and we need a shunting capacitance $C_\text{B} \simeq$ 63.9aF, so we would need a shunt capacitor much larger than the JJ. Note that the shunting capacitance is not a simple parallel plate capacitor, but it should take into account the entire capacitance matrix of the device.

In principle, the super-silicon JJ can be used to make a transmon qubit with similar design and geometry as the Al-based JJ. However, one must note that here the JJ critical current will be smaller than the Al-based JJ's, so physically it can't be any smaller for the same target frequency.

\subsection{Phase Slip Flux Qubit}

The high kinetic inductance and the possibility of making a very narrow wire makes the super-silicon a good candidate for the implementation of the phase slip qubit. In very narrow SC nanowires, fluctuations of the SC order parameter can be significant. Of particular interest here is when the amplitude of the SC order parameter reduces to near zero in a region of size $\xi_\mathrm{GL}$ allowing the phase to change by 2$\pi$. This phase slip can be considered as a macroscopic tunneling in the parameter space between two SC states with phase difference $\Delta\phi$=$2\pi$. This is responsible for the wide tail of resistance in SC transition in thin wires well below the critical temperature \cite{langer_ambegaokar_pr1967,mccumber_halperin_prb1970}. At very low temperatures, this macroscopic tunneling can still occur due to quantum fluctuations \cite{bezryadin_lau_nature2000,lau_markovic_prl2001,zaikin_golubev_prl1997,golubev_zaikin_prb2001,arutyunov_golubev_pr2008}. This quantum phase slip (QPS) is dual to the Cooper pair tunneling through a JJ \cite{mooij_nazarov_nphys2006,kerman_njp2013}, and devices dual to JJ devices were demonstrated \cite{hriscu_nazarov_prl2011,hriscu_nazarov_prb2011,hongisto_zorin_prl2012,lehtinen_zakharov_prl2012}. A new type of qubit \cite{mooij_harmans_njp2005}, called the phase slip (PS) flux qubit, was proposed in a SC loop with a QPS junction, where coherent QPS connects two states with different fluxoid quantum numbers. It is exactly dual to the charge qubit in CPB geometry, and the coherent superposition of qubit states was experimentally demonstrated \cite{astafiev_ioffe_nature2012,peltonen_astafiev_prb2013}. Full qubit operations have not yet been demonstrated.

\begin{figure}[!t]
  \centering
  \includegraphics[width=\linewidth]{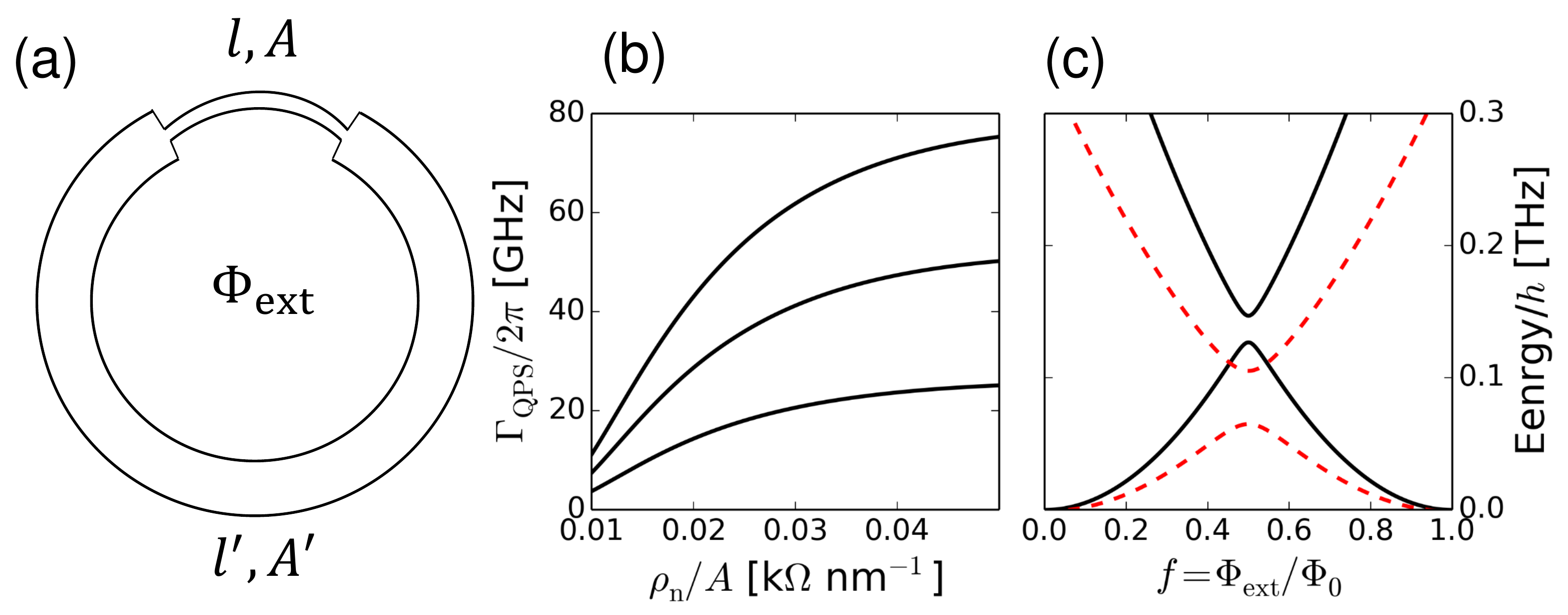}\\
  \caption{Phase slip flux qubit. 
           (a) A schematic of a phase slip flux qubit. The thin section of length $l$ and cross section area $A$ enables the coupling with different flux states via quantum phase slip. The remaining loop is much thicker and has a length of $l'$ and a cross section area $A'$. (b) Phase slip rate $\Gamma_{\text{QPS}}$ in the phase slip junction as a function of $\rho_\text{n}/A$ for a superconducting silicon device as described in the text, assuming $a$=$b$=1 in (\ref{eq:PSrate}). From bottom to top, the length $l$=100, 200, 300nm. (c) Energy spectrum for a loop with $l'$=1000nm, $A'$=1000nm$^2$, and $A$=64nm$^2$. The black solid curves are for $l$=100nm and the red dashed ones are for $l$=200nm.  }
  \label{fig:phaseslip}
\end{figure}

A schematic diagram of a PS flux qubit is in Fig. \ref{fig:phaseslip} (a). The thin section allows the phase slip to occur. Following the approaches in Refs. \cite{lau_markovic_prl2001} and \cite{mooij_harmans_njp2005}, the quantum phase slip rate is given by 
\begin{equation}\label{eq:PSrate}
\Gamma_{\text{QPS}} = 1.5 a \frac{l}{\xi_\text{GL}(0)} \sqrt{\frac{R_\text{Q}}{R_{\xi}}} \frac{k_\text{B} T_\text{c}}{\hbar} \exp\left( -0.3 b \frac{R_\text{Q}}{R_{\xi}} \right),
\end{equation}
where $a,b$ are fitting constants of order unity and $R_{\xi}$=$\rho_{n} \xi_\text{GL}/A$ is the normal resistance for a wire of length $\xi_\text{GL}$. In Fig. ~\ref{fig:phaseslip} (b), the phase slip rate is plotted as a function of $\rho_\text{n}/A$ for different lengths, $l$=100, 200, 300 nm. The target range of the rate would be of the order of a few GHz. For a phase slip junction with $l$=100 nm, we need $A \simeq$ 64nm$^2$ for 10GHz. For larger cross section areas, longer junctions would be needed to get the same phase slip rate.

The Hamiltonian for the PS qubit is
\begin{equation}\label{eq:Hphaseslip}
H_{\text{QPS}} = E_\text{L} ( f - \hat{N} )^2 + \frac{E_\text{S}}{2} \left( |N\rangle \langle N+1| + |N+1\rangle \langle N| \right),
\end{equation}
where $N$ is the fluxoid quantum number, and $E_L$=$\Phi_0^2/2L_K^{\text{tot}}$, and $E_s/2$=$\hbar \Gamma_{\text{QPS}}$.
$L_\text{K}^{\text{tot}}$ is the total kinetic inductance:
\begin{equation}
L_\text{K}^{\text{tot}} = L_\text{K} + L_\text{K}' = \frac{\hbar}{\pi} \frac{\rho_\text{n}}{\Delta_0} \left(  \frac{l}{A} + \frac{l'}{A'} \right).
\end{equation} 
The kinetic inductance is typically a few nH, while the geometric inductance is in pH range for the geometries considered. So we neglected geometric inductance here. The energy spectrum of the Hamiltonian in ({\ref{eq:Hphaseslip}) around a half-integer external flux is given in Fig.~\ref{fig:phaseslip} (c), for two different lengths $l$=100 and 200 nm. 

Flux is the relevant variable for the PS qubit, and the flux can be measured by a SQUID coupled to the phase qubit loop. However, near the optimal value of the external flux $f$=1/2 where the qubit states are insensitive to the flux noise up to the first order, the two eigenstates are linear combination of the two different flux states and can't be measured by directly measuring the flux. Instead we can use a resonator such as a lumped element LC resonator or a transmission line coupled to the qubit and perform the dispersive readout \cite{ilichev_prl2003,lupascu_prl2004}. As can be seen from Fig.~\ref{fig:phaseslip} (c), for longer $l$, phase slip energy $E_\text{S}$  increases while inductance energy $E_\text{L}$ decreases, which leads to flatter energy spectrum near the operation  point $f$=1/2. This is analogous to the charge qubit moving towards the transmon regime. Due to the exact duality with the charge qubit by a transformation~\cite{mooij_nazarov_nphys2006},
\begin{equation}
(\hat{q},\hat{\phi}) \leftrightarrow (-\hat{\phi/}2\pi, 2\pi \hat{q}), E_\text{S} \leftrightarrow E_\text{J},  E_\text{L} \leftrightarrow E_\text{C},
\end{equation}   
the transmon regime $E_\text{J}/E_\text{C} \gg 1$ corresponds to $E_\text{S}/E_\text{L} \gg 1$. This regime can be achieved, e.g. by increasing the total inductace while keeping $\Gamma_{\text{QPS}}$ constant.  

PS qubit has a few advantages over conventional SC qubits. A flux qubit implemented with a simple rf-SQUID loop needs very high loop and JJ inductances to form a well defined bound states, but this will suppress the tunneling between different flux states. Therefore, a widely adopted design for a JJ flux qubit is a SC loop with three JJs \cite{mooij_science1999,chiorescu_science2003}. PS qubit does not have this problem, and a simple loop with a single PS junction is enough. The excited states have a much higher energy than the qubit states compared to the transmon qubit, so much faster gate operations are possible without exciting the qubit out of the qubit subspace (leakage). It is insensitive to the charge noise since there is no SC island, and the phase slip rate is determined by the geometry of the thin wire on the scale of the coherence length, making it insensitive to microscopic fluctuations of individual atoms \cite{mooij_harmans_njp2005}. A main source of decoherence could be the flux noise, but we expect it could be less of a problem for the PS qubit embedded in a pure silicon crystal. Also, we can increase the total inductance to move into the transmon-like regime to suppress the flux noise, at the cost of reduced anharmonicity.  In the context of super-silicon devices, the simple washer-like geometry eases fabrication requirements and the need for a precisely controlled tunnel gap.

\section{Applications for KI Detectors}

\begin{figure}[!t]
  \centering
  \includegraphics[width=\linewidth]{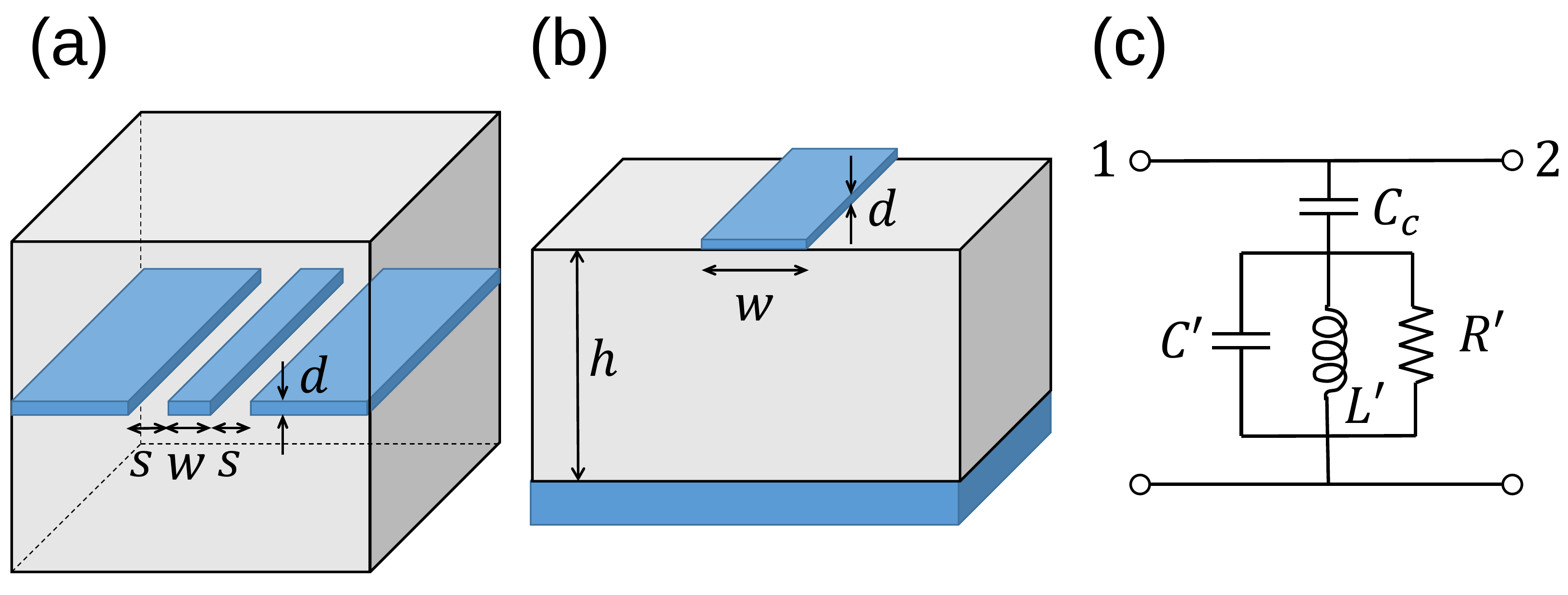}\\
  \caption{MKID particle detector. (a) Coplanar waveguide (CPW) transmission line. The SC resonator is made of a $\lambda/4$ CPW. The center strip of width $w$ is separated from the ground planes by a gap $s$. The whole device can be embedded inside a single silicon crystal to potentially reduce loss. The superconducting gap ($T_\text{c}$) can be tuned by tuning the density of superconducting semiconductor region. Superconducting-semiconductors offer natural high kinetic inductance as calculated in the text. (b) Microstrip transmission line. The bottom SC layer is the ground plane, and the top microstrip is patterned to make a transmission line resonator. The ground plane can be formed by the superconducting-semiconductor region with a high-quality crystalline over-growth acting as the gap. (c) Schematic circuit diagram for the device. A transmission line is coupled to the resonator. The change in the resonance frequency and the quality factor can be detected by the transmission coefficient $t_{21}$ between port 1 and port 2. }
  \label{fig:MKID}
\end{figure}

Finally we consider the microwave kinetic inductance particle detector (MKID) devices \cite{day_nature2003,zmuidzinas_review}. If a photon (or phonon) with energy larger than the SC gap is absorbed into a SC resonator, it breaks Cooper pairs and creates quasi-particles. This increases the residual resistance and the kinetic inductance and therefore changes the resonator quality factor and the resonance frequency. This change can be detected by measuring the transmitted signal through a nearby transmission line capacitively coupled to the resonator. A schematic circuit diagram for an MKID detector is shown in Fig.~\ref{fig:MKID}(c).  This scheme allows easy frequency division multiplexing and and it can be deployed in a large array. MKID is currently being developed for a large range of frequencies, from IR to X-ray photons, and especially for astronomy applications. 

The coplanar waveguide (CPW) [see Fig. \ref{fig:MKID}(a)] has become the preferred choice for the resonator design \cite{zmuidzinas_review}. Given the properties of super-silicon, we will consider a CPW resonator in the thin film and local limit ($w \ll \lambda$). The thin film CPW resonator of length $l$ has a surface impedance $Z_\text{s}$=$R_\text{s}$+$i X_\text{s}$ where $R_\text{s}$ is the surface resistance due to the thermally excited or photon-created quasi-particles, and $X_\text{s}$=$\omega L_\text{s}$ with the surface inductance $L_\text{s}$ being related to the kinetic inductance $L_\text{K}$=$L_\text{S} l/w$. The resonator has an internal quality factor $Q_\text{i}$ and the coupling quality factor $Q_\text{c}$ due to the coupling to the transmission line. The total resonator Q-factor $Q_\text{r}$ is $1/Q_\text{r}$=$1/Q_\text{i}+1/Q_\text{c}$. $Q_\text{c}$ can be tuned by changing the coupling capacitance $C_\text{c}$. The resonator has distributed geometrical capacitance $C'$, geometrical inductance $L'$, and resistance $R'$. Geometrical capacitance and inductance of a CPW resonator can be calculated using analytical expressions derived by conformal mapping \cite{gao_thesis}. The total inductance is the sum of geometrical and kinetic inductances. For a short-ended quarter-wave resonator, the change in the transmission coefficient near the resonance frequency is given by \cite{gao_thesis}
\begin{equation}
\delta t_{21} \mid_{f=f_\text{r}} = \frac{\delta Q_\text{r}^2}{Q_\text{c}} \left[\delta\left( \frac{1}{Q_\text{i}} \right) - 2i\frac{\delta f_\text{r}}{f_\text{r}} \right] \simeq \alpha \frac{\delta Q_\text{r}^2}{Q_\text{c}} \frac{\delta Z_\text{s}}{|Z_\text{s}|} ~, 
\end{equation} 
where $\alpha$ is the fraction of kinetic inductance and $f_\text{r}$=$1/4l\sqrt{L'C'}$ is the resonance frequency of the CPW resonator. $Q_\text{r}^2/Q_\text{c}$ is maximized if $Q_\text{c}$ is set to be equal to $Q_\text{i}$. $\alpha$ is larger for smaller resonators, i.e. thinner and narrower CPW resonator has a higher fraction of kinetic inductance.

\begin{figure}[!t]
  \centering
  \includegraphics[width=\linewidth]{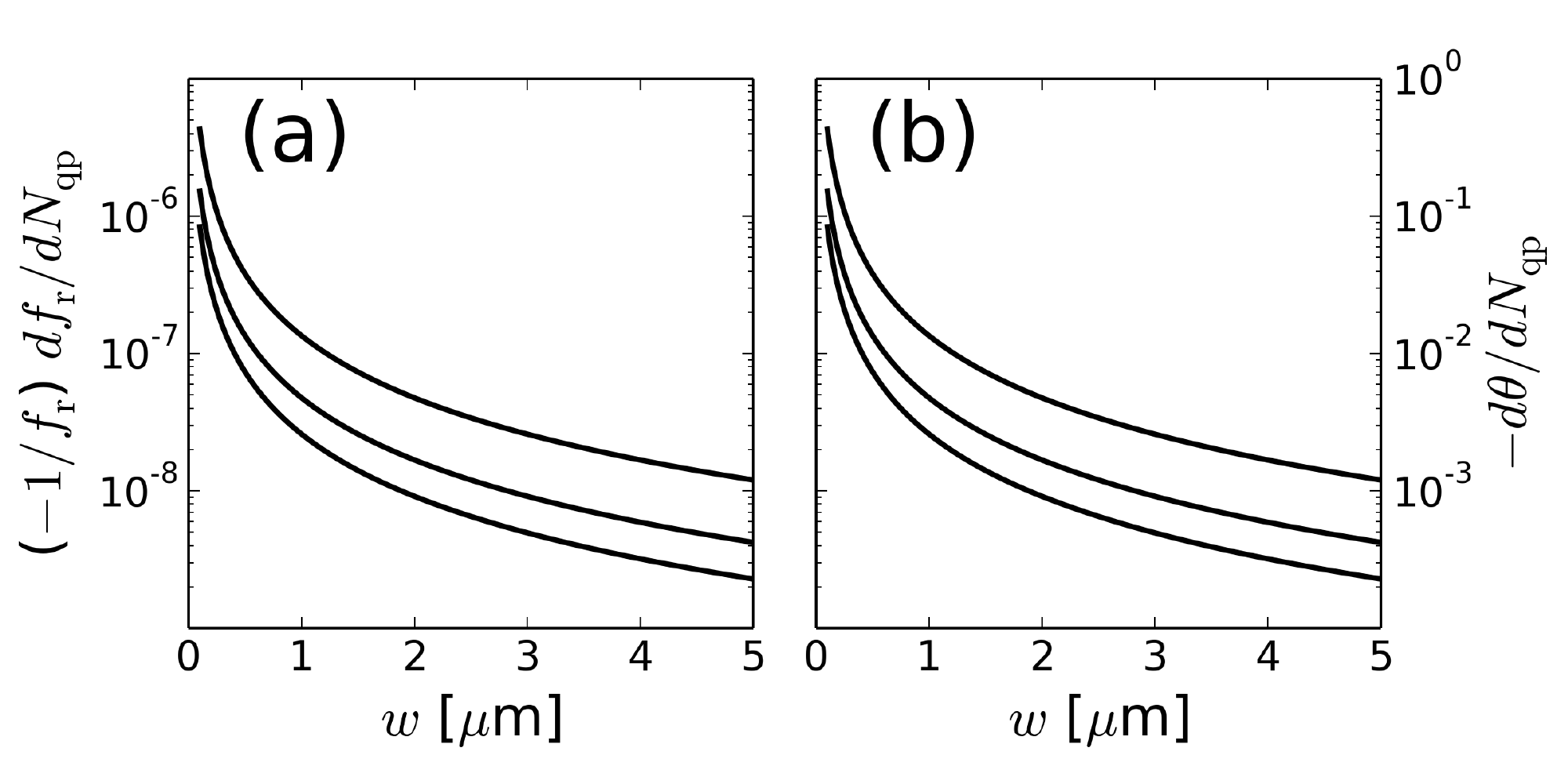}\\
  \caption{Efficiency of MKID devices made of super-silicon, as a function of the width $w$ of the center strip. $T$=10mK and the ratio $w/s$=2/3 was fixed. $Q_\text{i}$=$10^5$ was assumed. (a) Change in the resonance frequency. (b) Responsivity of MKID, which is the rate of change in the phase of the transmission coefficient. From top to bottom, the depth $d$=10nm, 20nm, 30nm.}
  \label{fig:MKIDresponse}
\end{figure}

We calculated the shift in the resonance frequency and the responsivity of the MKID for super-silicon. Fig. \ref{fig:MKIDresponse} shows the change of resonance frequency [(a)] and the phase angle of the transmission coefficient [(b)] as quasi-particles are created by incident photons. The efficiency of the super-silicon MKID is expected to be quite good, compared to other metallic superconductors currently used in MKID devices \cite{leluc_bumble_apl2010,cecil_miceli_apl2012,gao_vissers_apl2012,calvo_daddabbo_jltp2014}. We assumed a internal quality factor of $Q_\text{i}$=10$^5$ which was achieved in other materials \cite{cecil_miceli_apl2012}. Due to the available semiconductor technologies for preparing very ideal, chemically purified, and even isotopically enriched semiconductor crystals, we expect even higher $Q_\text{i}$ could be possible. The STM lithography technique could enable much smaller devices than current MKID devices, which could also enhance the sensitivity of the detector.

The historical path of MKID devices has progressed toward CPW structures as they are easier to fabricate, requiring only a single layer of SC metal, and have shown very high Q's, especially in the last decade. However, one of the originally considered geometries, the microstrip resonator [shown in \ref{fig:MKID}(b)], may be of interest in the context of SC semiconductor devices. It consists of a ground plane at the bottom, a substrate separation layer with height $h$, and a transmission line resonator on top. Although more complicated, the theoretical kinetic inductance fraction can approach unity in such structures while simultaneously having a small active volume (though losses in the dielectric separation layer or the thickness of the separation layer have limited performance in the past)  \cite{zmuidzinas_review}.  In the super-semi approach, the separation layer could be fully epitaxial and quite thin. And, practically, the bottom ground plane does not require nanoscale precision (STM lithography): the SC region can be doped by other 2D methods, such as gas immersion laser doping (GILD) \cite{bustarret_nature2006}.

If we need larger SC energy gap to detect higher energy photons, boron-doped diamond might be suitable. It has much higher critical temperature ($\sim$ 7K) and larger normal resistivity than boron-doped silicon (see Table.\ref{tab:kinetic_inductance}). The kinetic inductance is smaller but comparable. SC diamond is also a type II BCS superconductor in the dirty limit like super-silicon, and clean single crystal SC diamond epilayers have been successfully grown and the vortex lattice structure was observed \cite{bustarret_achatz_ptrsa2008}. 

Most interestingly, the super-semi approach allows one to tune the superconducting gap, $\Delta$=1.76 $k_\text{B} T_\text{c}$, by changing the density of holes in the superconducting region since the critical temperature $T_\text{c} \propto (n_\text{B}/n_\text{c}-1)^{1/2}$ as experimentally observed for C:B \cite{bustarret_achatz_ptrsa2008}, Si:B \cite{marcenat_prb2010}, and Ge:Ga \cite{skrotzki_ltp2011} where $c_\text{B}$ is the doped acceptor density and $c_\text{c}$ is the critical acceptor density for SC transition. This way, the device can be directly engineered for specific target particle energies or for quasiparticle trapping or movement.

These type of detectors are also being used in the search for weakly interacting massive particles (WIMPs) which have been proposed to comprise dark matter \cite{wimp_review}. The Cryogenic Dark Matter Search (CDMS) detectors use bulk silicon or germanium as a target to absorb WIMPs with MKID phonon detectors covering the top \cite{cdms_prl2011,cdms_prl2013}. MKIDs can be tuned to detect the phonons created by absorbed WIMPs \cite{daal_jltp2008,golwala_jltp2008}. Super-silicon or germanium seem to be an attractive material for this since the MKID detector and the target are made of the same material and the SC energy gap can be easily tuned by tuning the hole density in the SC regions. 

Phonon detectors may also have direct relevance to the rapidly growing field of opto-/nanomechanics \cite{ekinci_roukes_nems_review,schwab_roukes_phystoday2005,kippenberg_vahala_science2008,marquardt_girvin_physics2009}. It's interesting to estimate the possibility of use of super-semi MKIDs for acoustic phonon detection. Here, one would want the $T_\text{c}$ sufficiently high to limit the thermal noise due to thermal activation of quasi-particles, while having $T_\text{c}$ low enough to allow for acoustic phonons to create quasiparticles that are then detected. Taking as an example phonons of 10 GHz that one would wish to detect in a silicon device, one would want to engineer $T_\text{c} \approx$ 0.12 K such that $2\Delta_0 \lesssim$ 10 GHz. This would correspond to a hole density of roughly 2 - 3 $\times 10^{21}$ cm$^{-3}$. Signal noise to sensitivity would improve with increasing phonon energy.

\section{Conclusions}
We have derived the characteristic superconducting parameters for heavily boron-doped silicon as well as relevant JJ parameters for precision-doped super-silicon devices. Applications of this system for qubits (the transmon qubit and the phase-slip qubit) and for MKID particle detectors have been proposed, and possible device designs were suggested. STM hydrogen lithography-enabled acceptor doping is not yet realized, but it appears that there is no fundamental reason against its implementation in the near future. 
Once realized, this bottom-up fabrication technique offers extreme flexibility in device design down to the single lattice-site level.  Due to the flexibility of material parameters (hole density, $T_\text{c}$, $\rho_\text{n}$, etc.), different components (superconducting regions, normal metal regions, insulators) could all be brought together inside a nearly-perfect epitaxial crystal. Although we presumed (STM) hydrogen lithography as a preferable technology for precision impurity placement to make super-silicon devices, examples proposed in this work could be realized in SC semiconductors that are already available through other methods (which do not require nanoscale control), examples include already demonstrated doping techniques for superconducting silicon, germanium, and diamond. Germanium in particular could be especially amenable to three-dimensional atomic-layer doping. Superconducting-diamond offers high-T$_c$, a large kinetic inductance as shown above, and compatibility with many interesting device applications from nanomechanics to photonics.
It may be also worth trying different types of acceptors such as aluminum or gallium both in silicon and germanium for better material properties \cite{bourgeois_apl2007}. There seems to be no fundamental reasons preventing superconductivity in electron-doped semiconductors \cite{cohen_rmp1964,bourgeois_apl2007}, but there has been no experimental evidence yet \cite{grockowiak_sst2013}.

If atomic-layer doping is available, but precision hydrogen lithography is not, it should be possible to make a parallel-plate transmon device of Section V.A above by growing two superconducting layers epitaxially separated by a distance $d$, then etching devices of area $A$ for a given target frequency using standard lithographic techniques. Although in this case there are clearly interfaces near the device, the electric field could be localized in the epitaxial region with possibly improved device performance. This double layer device might also be created via the gas immersion laser doping technique \cite{bustarret_nature2006}. Single layer superconducting devices such as the phase-slip qubit are simpler, requiring only a single superconducting region followed by patterned etching.
We have estimated the kinetic inductance of these materials and have shown how they could be used as particle detectors, perhaps in the very near term. The true potential of superconducting semiconductor devices, as discussed only specifically here, lies in the possibility of completely new device designs and of a testbed for new science due to the flexibility of the bottom-up fabrication and the huge physics-space of SC circuits and systems.


\section*{Acknowledgment}
The authors thank B. Palmer, R. Schoelkopf, and M. Devoret for useful conversations.


\bibliographystyle{IEEEtran}

\begin{thebibliography}{71}
\providecommand{\url}[1]{#1}
\csname url@samestyle\endcsname
\providecommand{\newblock}{\relax}
\providecommand{\bibinfo}[2]{#2}
\providecommand{\BIBentrySTDinterwordspacing}{\spaceskip=0pt\relax}
\providecommand{\BIBentryALTinterwordstretchfactor}{4}
\providecommand{\BIBentryALTinterwordspacing}{\spaceskip=\fontdimen2\font plus
\BIBentryALTinterwordstretchfactor\fontdimen3\font minus
  \fontdimen4\font\relax}
\providecommand{\BIBforeignlanguage}[2]{{%
\expandafter\ifx\csname l@#1\endcsname\relax
\typeout{** WARNING: IEEEtran.bst: No hyphenation pattern has been}%
\typeout{** loaded for the language `#1'. Using the pattern for}%
\typeout{** the default language instead.}%
\else
\language=\csname l@#1\endcsname
\fi
#2}}
\providecommand{\BIBdecl}{\relax}
\BIBdecl

\bibitem{cohen_rmp1964}
M.~L. Cohen, ``The existence of a superconducting state in semiconductors,''
  \emph{Rev. Mod. Phys.}, vol.~36, pp. 240--243, 1964.

\bibitem{ekimov_nature2004}
E.~A. Ekimov, V.~A. Sidorov, E.~D. Bauer, N.~N. Mel'nik, N.~J. Curro, J.~D.
  Thompson, and S.~M. Stishov, ``Superconductivity in diamond,'' \emph{Nature},
  vol. 428, pp. 542--545, 2004.

\bibitem{bustarret_nature2006}
E.~Bustarret, C.~Marcenat, P.~Achatz, J.~Kacmarcik, F.~L{\'e}vy, A.~Huxley,
  L.~Ort{\'e}ga, E.~Bourgeois, X.~Blas{\'e}, D.~D{\'e}barre, and J.~Boulmer,
  ``Superconductivity in doped cubic silicon,'' \emph{Nature}, vol. 444, pp.
  465--468, 2006.

\bibitem{herrmannsdorfer_prl2009}
T.~Herrmannsd{\"o}rfer, V.~Heera, O.~Ignatchik, M.~Uhlarz, A.~M{\"u}cklich,
  M.~Posselt, H.~Reuther, B.~Schmidt, K.-H. Heinig, W.~Skorupa, M.~Voelskow,
  C.~W{\"u}ndisch, R.~Skrotzki, M.~Helm, and J.~Wosnitza, ``Superconducting
  state in a gallium-doped germanium layer at low temperatures,'' \emph{Phys.
  Rev. Lett.}, vol. 102, p. 217003, 2009.

\bibitem{blase_nmat2009}
X.~Blase, E.~Bustarret, C.~Chapelier, T.~Klein, and C.~Marcenat,
  ``superconducting group-iv semiconductors,'' \emph{Nat. Mater.}, vol.~8, pp.
  375--382, 2009.

\bibitem{fuhrer_nanolett2009}
A.~Fuhrer, M.~F{\"u}chsle, T.~C.~G. Reusch, B.~Weber, and M.~Y. Simmons,
  ``Atomic-scale, all epitaxial in-plane gated donor quantum dot in silicon,''
  \emph{Nano Lett.}, vol.~9, pp. 707--710, 2009.

\bibitem{fuechsle_nnano2010}
M.~Fuechsle, S.~Mahapatra, F.~A. Zwanenburg, M.~Friesen, M.~A. Eriksson, and
  M.~Y. Simmons, ``Spectroscopy of few-electron single-crystal silicon quantum
  dots,'' \emph{Nat. Nanotechnol.}, vol.~5, pp. 502--505, 2010.

\bibitem{shim_tahan_ncomm2014}
Y.-P. Shim and C.~Tahan, ``Bottom-up superconducting and josephson junction
  device inside a group-iv semiconductor,'' arXiv:1309.0015.

\bibitem{SC_qubit_review}
M.~H. Devoret and R.~J. Schoelkopf, ``Superconducting circuits for quantum
  information: An outlook,'' \emph{Science}, vol. 339, pp. 1169--1174, 2013.

\bibitem{blais_pra2004}
A.~Blais, R.-S. Huang, A.~Wallraff, S.~M. Girvin, and R.~J. Schoelkopf,
  ``Cavity quantum electrodynamics for superconducting electrical circuits: An
  architecture for quantum computation,'' \emph{Phys. Rev. A}, vol.~69, p.
  062320, 2003.

\bibitem{transmon}
J.~Koch, T.~M. Yu, J.~Gambetta, A.~A. Houck, D.~I. Schuster, J.~Majer,
  A.~Blais, M.~H. Devoret, S.~M. Girvin, and R.~J. Schoelkopf,
  ``Charge-insensitive qubit design derived from the cooper pair box,''
  \emph{Phys. Rev. A}, vol.~76, p. 042319, 2007.

\bibitem{paik_prl2011}
H.~Paik, D.~I. Schuster, L.~S. Bishop, G.~Kirchmair, G.~Catelani, A.~P. Sears,
  B.~R. Johnson, M.~J. Reagor, L.~Frunzio, L.~I. Glazman, S.~M. Girvin, M.~H.
  Devoret, and R.~J. Schoelkopf, ``Observation of high coherence in josephson
  junction qubits measured in a three-dimensional circuit qed architecture,''
  \emph{Phys. Rev. Lett.}, vol. 107, p. 240501, 2011.

\bibitem{oh_prb2006}
S.~Oh, K.~Cicak, J.~S. Kline, M.~A. Sillanp{\"aa}, K.~D. Osborn, J.~D.
  Whittaker, R.~W. Simmonds, and D.~P. Pappas, ``Elimination of two level
  fluctuators in superconducting quantum bits by an epitaxial tunnel barrier,''
  \emph{Phys. Rev. B}, vol.~74, p. 100502(R), 2006.

\bibitem{pappas1}
J.~S. Kline, M.~R. Vissers, F.~C.~S. da~Silva, D.~S. Wisbey, M.~Weides, T.~J.
  Weir, B.~Turek, D.~A. Braje, W.~D. Oliver, Y.~Shalibo, N.~Katz, B.~R.
  Johnson, T.~A. Ohki, and D.~P. Pappas, ``Sub-micrometer epitaxial josephson
  junctions for quantum circuits,'' \emph{Supercond. Sci. Technol.}, vol.~25,
  p. 025005, 2012.

\bibitem{pappas2}
J.~B. Chang, M.~R. Vissers, A.~D. Corcoles, M.~Sandberg, J.~Gao, D.~W. Abraham,
  J.~M. Chow, J.~M. Gambetta, M.~B. Rothwell, G.~A. Keefe, M.~Steffen, and
  D.~P. Pappas, ``Improved superconducting qubit coherence using titanium
  nitride,'' \emph{Appl. Phys. Lett.}, vol. 103, p. 012602, 2013.

\bibitem{sendelbach_prl2009}
S.~Sendelbach, D.~Hover, M.~Mu{\" c}k, and R.~McDermott, ``Complex inductance,
  excess noise, and surface magnetism in dc squids,'' \emph{Phys. Rev. Lett.},
  vol. 103, p. 117001, 2009.

\bibitem{fuechsle_nnano2012}
M.~Fuechsle, J.~A. Miwa, S.~Mahapatra, H.~Ryu, S.~Lee, O.~Warschkow, L.~C.~L.
  Hollenberg, G.~Klimeck, and M.~Y. Simmons, ``A single-atom transistor,''
  \emph{Nat. Nanotechnol.}, vol.~7, pp. 242--246, 2012.

\bibitem{weber_science2012}
B.~Weber, S.~Mahapatra, H.~Ryu, S.~Lee, A.~Fuhrer, T.~C.~G. Reusch, D.~L.
  Thompson, W.~C.~T. Lee, G.~Klimeck, L.~C.~L. Hollenberg, and M.~Y. Simmons,
  ``Ohm's law survives to the atomic scale,'' \emph{Science}, vol. 335, pp.
  64--67, 2012.

\bibitem{rusko_prb2013}
R.~Ruskov and C.~Tahan, ``On-chip cavity quantum phonodynamics with an acceptor
  qubit in silicon,'' \emph{Phys. Rev. B}, vol.~88, p. 064308, 2013.

\bibitem{mooij_harmans_njp2005}
J.~E. Mooij and C.~J. P.~M. Harmans, ``Phase-slip flux qubits,'' \emph{New J.
  Phys.}, vol.~7, p. 219, 2005.

\bibitem{BCS}
J.~Bardeen, L.~N. Cooper, and J.~R. Schrieffer, ``Theory of
  superconductivity,'' \emph{Phys. Rev.}, vol. 108, pp. 1175--1204, 1957.

\bibitem{marcenat_prb2010}
C.~Marcenat, J.~Ka{\v c}mar{\v c}{\'i}k, R.~Piquerel, P.~Achatz, G.~Prudon,
  C.~Dubois, B.~Gautier, J.~C. Dupuy, E.~Bustarret, L.~Ortega, T.~Klein,
  J.~Boulmer, T.~Kociniewski, and D.~D{\'e}barre, ``Low-temperature transition
  to a superconducting phase in boron-doped silicon films grown on
  (001)-oriented silicon wafers,'' \emph{Phys. Rev. B}, vol.~81, p. 020501(R),
  2010.

\bibitem{bourgeois_apl2007}
E.~Bourgeois and X.~Blase, ``Superconductivity in doped cubic silicon: An {\it
  ab initio} study,'' \emph{Appl. Phys. Lett.}, vol.~90, p. 142511, 2007.

\bibitem{ambegaokar_prl1963}
V.~Ambegaokar and A.~Baratoff, ``Tunneling between superconductors,''
  \emph{Phys. Rev. Lett.}, vol.~10, pp. 486--489, 1963.

\bibitem{annunziata_nanotech2010}
A.~J. Annunziata, D.~F. Santavicca, L.~Frunzio, G.~Catelani, M.~J. Rooks,
  A.~Frydman, and D.~E. Prober, ``Tunable superconducting nanoinductors,''
  \emph{Nanotechnology}, vol.~21, p. 445202, 2010.

\bibitem{Mattis_Bardeen}
D.~C. Mattis and J.~Bardeen, ``Theory of the anomalous skin effect in normal
  and superconducting metals,'' \emph{Phys. Rev.}, vol. 111, pp. 412--417,
  1958.

\bibitem{McCambridge_dissert}
J.~D. McCambridge, ``The superconducting properties of niobium-titanium alloy
  multilayers,'' Ph.D. dissertation, Yale University, 1995.

\bibitem{gao_hajenius_apl2007}
J.~R. Gao, M.~Hajenius, F.~D. Tichelaar, T.~M. Klapwijk, B.~Voronov,
  E.~Grishin, G.~Gol’tsman, C.~A. Zorman, and M.~Mehregany, ``Monocrystalline
  nbn nanofilms on a 3 c - si c si substrate,'' \emph{Appl. Phys. Lett.},
  vol.~91, p. 062504, 2007.

\bibitem{calvo_daddabbo_jltp2014}
M.~Calvo, A.~D’Addabbo, A.~Monfardini, A.~Benoit, N.~Boudou, O.~Bourrion,
  A.~Catalano, L.~Dumoulin, J.~Goupy, H.~L. Sueur, and S.~Marnieros, ``Niobium
  silicon alloys for kinetic inductance detectors,'' \emph{J. Low Temp. Phys.},
  2014, dOI 10.1007/s10909-013-1072-6.

\bibitem{leluc_bumble_apl2010}
H.~G. Leduc, B.~Bumble, P.~K. Day, B.~H. Eom, J.~Gao, S.~Golwala, B.~A. Mazin,
  S.~McHugh, A.~Merrill, D.~C. Moore, O.~Noroozian, A.~D. Turner, , and
  J.~Zmuidzinas, ``Titanium nitride films for ultrasensitive microresonator
  detectors,'' \emph{Appl. Phys. Lett.}, vol.~97, p. 102509, 2010.

\bibitem{barends_hortensius_apl2008}
R.~Barends, H.~L. Hortensius, T.~Zijlstra, J.~J.~A. Baselmans, S.~J.~C. Yates,
  J.~R. Gao, and T.~M. Klapwijk, ``Contribution of dielectrics to frequency and
  noise of nbtin superconducting resonators,'' \emph{Appl. Phys. Lett.},
  vol.~92, p. 223502, 2008.

\bibitem{ishizaka_prl2007}
K.~Ishizaka, R.~Eguchi, S.~Tsuda, T.~Yokoya, A.~Chainani, T.~Kiss,
  T.~Shimojima, T.~Togashi, S.~Watanabe, C.-T. Chen, C.~Q. Zhang, Y.~Takano,
  M.~Nagao, I.~Sakaguchi, T.~Takenouchi, H.~Kawarada, and S.~Shin,
  ``Observation of a superconducting gap in boron-doped diamond by
  laser-excited photoemission spectroscopy,'' \emph{Phys. Rev. Lett.}, vol.~98,
  p. 047003, 2007.

\bibitem{skrotzki_ltp2011}
R.~Skrotzki, T.~Herrmannsd{\"o}rfer, V.~Heera, J.~Fiedler, A.~M{\"u}cklich,
  M.~Helm, and J.~Wosnitza, ``The impact of heavy ga doping on
  superconductivity in germanium,'' \emph{Low Temp. Phys.}, vol.~37, pp.
  877--883, 2011.

\bibitem{transmon_exp}
J.~A. Schreier, A.~A. Houck, J.~Koch, D.~I. Schuster, B.~R. Johnson, J.~M.
  Chow, J.~M. Gambetta, J.~Majer, L.~Frunzio, M.~H. Devoret, S.~M. Girvin, ,
  and R.~J. Schoelkopf, ``Suppressing charge noise decoherence in
  superconducting charge qubits,'' \emph{Phys. Rev. B}, vol.~77, p. 180502(R),
  2008.

\bibitem{wallraff_nature2004}
A.~Wallraff, D.~I. Schuster, A.~Blais, L.~Frunzio, R.-S. Huang, J.~Majer,
  S.~Kumar, S.~M. Girvin, and R.~J. Schoelkopf, ``Strong coupling of a single
  photon to a superconducting qubit using circuit quantum electrodynamics,''
  \emph{Nature}, vol. 431, pp. 162--167, 2004.

\bibitem{langer_ambegaokar_pr1967}
J.~S. Langer and V.~Ambegaokar, ``Intrinsic resistive transition in narrow
  suyerconducting channels,'' \emph{Phys. Rev.}, vol. 164, p. 498, 1967.

\bibitem{mccumber_halperin_prb1970}
D.~E. McCumber and B.~I. Halperin, ``Time scale of intrinsic resistive
  fluctuations in thin superconducting wires,'' \emph{Phys. Rev. B}, vol.~1, p.
  1054, 1970.

\bibitem{bezryadin_lau_nature2000}
A.~Bezryadin, C.~N. Lau, and M.~Tinkham, ``Quantum suppression of
  superconductivity in ultrathin nanowires,'' \emph{Nature}, vol. 404, p. 971,
  2000.

\bibitem{lau_markovic_prl2001}
C.~N. Lau, N.~Markovic, M.~Bockrath, A.~Bezryadin, and M.~Tinkham, ``Quantum
  phase slips in superconducting nanowires,'' \emph{Phys. Rev. Lett.}, vol.~87,
  p. 217003, 2001.

\bibitem{zaikin_golubev_prl1997}
A.~D. Zaikin, D.~S. Golubev, A.~van Otterlo, , and G.~T. Zim{\' a}nyi,
  ``Quantum phase slips and transport in ultrathin superconducting wires,''
  \emph{Phys. Rev. Lett.}, vol.~78, p. 1552, 1997.

\bibitem{golubev_zaikin_prb2001}
D.~S. Golubev and A.~D. Zaikin, ``Quantum tunneling of the order parameter in
  superconducting nanowires,'' \emph{Phys. Rev. B}, vol.~64, p. 014504, 2001.

\bibitem{arutyunov_golubev_pr2008}
K.~Y. Arutyunov, D.~S. Golubev, and A.~D. Zaikin, ``Superconductivity in one
  dimension,'' \emph{Phys. Rep.}, vol. 464, p.~1, 2008.

\bibitem{mooij_nazarov_nphys2006}
J.~E. Mooij and Y.~V. Nazarov, ``Superconducting nanowires as quantum
  phase-slip junctions,'' \emph{Nat. Phys.}, vol.~2, pp. 169--172, 2006.

\bibitem{kerman_njp2013}
A.~J. Kerman, ``Flux–charge duality and topological quantum phase
  fluctuations in quasi-one-dimensional superconductors,'' \emph{New J. Phys.},
  vol.~15, p. 105017, 2013.

\bibitem{hriscu_nazarov_prl2011}
A.~M. Hriscu and Y.~V. Nazarov, ``Model of a proposed superconducting phase
  slip oscillator: A method for obtaining few-photon nonlinearities,''
  \emph{Phys. Rev. Lett.}, vol. 106, p. 077004, 2011.

\bibitem{hriscu_nazarov_prb2011}
------, ``Coulomb blockade due to quantum phase slips illustrated with
  devices,'' \emph{Phys. Rev. B}, vol.~83, p. 174511, 2011.

\bibitem{hongisto_zorin_prl2012}
T.~T. Hongisto and A.~B. Zorin, ``Single-charge transistor based on the
  charge-phase duality of a superconducting nanowire circuit,'' \emph{Phys.
  Rev. Lett.}, vol. 108, p. 097001, 2012.

\bibitem{lehtinen_zakharov_prl2012}
J.~S. Lehtinen, K.~Zakharov, and K.~Y. Arutyunov, ``Coulomb blockade and bloch
  oscillations in superconducting ti nanowires,'' \emph{Phys. Rev. Lett.}, vol.
  109, p. 187001, 2012.

\bibitem{astafiev_ioffe_nature2012}
O.~V. Astafiev, L.~B. Ioffe, S.~Kafanov, Y.~A. Pashkin, K.~Y. Arutyunov,
  D.~Shahar, O.~Cohen, and J.~S. Tsai, ``Coherent quantum phase slip,''
  \emph{Nature}, vol. 484, pp. 355--358, 2012.

\bibitem{peltonen_astafiev_prb2013}
J.~T. Peltonen, O.~V. Astafiev, Y.~P. Korneeva, B.~M. Voronov, A.~A. Korneev,
  I.~M. Charaev, A.~V. Semenov, G.~N. Golt’sman, L.~B. Ioffe, T.~M. Klapwijk,
  and J.~S. Tsai, ``Coherent flux tunneling through nbn nanowires,''
  \emph{Phys. Rev. B}, vol.~88, p. 220506(R), 2013.

\bibitem{ilichev_prl2003}
E.~Il'ichev, N.~Oukhanski, A.~Izmalkov, T.~Wagner, M.~Grajcar, H.-G. Meyer,
  A.~Y. Smirnov, A.~M. van~den Brink, M.~H.~S. Amin, and A.~M. Zagoskin,
  ``Continuous monitoring of rabi oscillations in a josephson flux qubit,''
  \emph{Phys. Rev. Lett.}, vol.~91, p. 097906, 2003.

\bibitem{lupascu_prl2004}
A.~Lupas{\c c}u, C.~J.~M. Verwijs, R.~N. Schouten, C.~J. P.~M. Harmans, and
  J.~E. Mooij, ``Nondestructive readout for a superconducting flux qubit,''
  \emph{Phys. Rev. Lett.}, vol.~93, p. 177006, 2004.

\bibitem{mooij_science1999}
J.~E. Mooij, T.~P. Orlando, L.~Levitov, L.~Tian, C.~H. van~der Wal, and
  S.~Lloyd, ``Josephson persistent-current qubit,'' \emph{Science}, vol. 285,
  pp. 1036--1039, 1999.

\bibitem{chiorescu_science2003}
I.~Chiorescu, Y.~Nakamura, C.~J. P.~M. Harmans, and J.~E. Mooij, ``Coherent
  quantum dynamics of a superconducting flux qubit,'' \emph{Science}, vol. 299,
  pp. 1869--1871, 2003.

\bibitem{day_nature2003}
P.~K. Day, H.~G. LeDuc, B.~A. Mazin, A.~Vayonakis, and J.~Zmuidzinas, ``A
  broadband superconducting detector suitable for use in large arrays,''
  \emph{Nature}, vol. 425, pp. 817--821, 2003.

\bibitem{zmuidzinas_review}
J.~Zmuidzinas, ``Superconducting microresonators: Physics and applications,''
  \emph{Annu. Rev. Condens. Matter Phys.}, vol.~3, pp. 169--214, 2012.

\bibitem{gao_thesis}
J.~Gao, ``The physics of superconducting microwave resonators,'' Ph.D.
  dissertation, California Institute of Technology, 2008.

\bibitem{cecil_miceli_apl2012}
T.~Cecil, A.~Miceli, O.~Quaranta, C.~Liu, D.~Rosenmann, S.~McHugh, and
  B.~Mazin, ``Tungsten silicide films for microwave kinetic inductance
  detectors,'' \emph{Appl. Phys. Lett.}, vol. 101, p. 032601, 2012.

\bibitem{gao_vissers_apl2012}
J.~Gao, M.~R. Vissers, M.~O. Sandberg, F.~C.~S. da~Silva, S.~W. Nam, D.~P.
  Pappas, D.~S. Wisbey, E.~C. Langman, S.~R. Meeker, B.~A. Mazin, H.~G. Leduc,
  J.~Zmuidzinas, and K.~D. Irwin, ``A titanium-nitride near-infrared kinetic
  inductance photon-counting detector and its anomalous electrodynamics,''
  \emph{Appl. Phys. Lett.}, vol. 101, p. 142602, 2012.

\bibitem{bustarret_achatz_ptrsa2008}
E.~Bustarret, P.~Achatz, B.~Sac{\'e}p{\'e}, C.~Chapelier, C.~Marcenat,
  L.~Ort{\'e}ga, and T.~Klein, ``Metal-to-insulator transition and
  superconductivity in boron-doped diamond,'' \emph{Phil. Trans. R. Soc. A},
  vol. 366, pp. 267--279, 2008.

\bibitem{wimp_review}
G.~Jungman, M.~Kamionkowski, and K.~Griest, ``Supersymmetric dark matter,''
  \emph{Phys. Rep.}, vol. 267, pp. 195--373, 1996.

\bibitem{cdms_prl2011}
``Results from a low-energy analysis of the cdms ii germanium data,''
  \emph{Phys. Rev. Lett.}, vol. 106, p. 131302, Mar 2011.

\bibitem{cdms_prl2013}
``Silicon detector dark matter results from the final exposure of cdms ii,''
  \emph{Phys. Rev. Lett.}, vol. 111, p. 251301, Dec 2013.

\bibitem{daal_jltp2008}
M.~Daal, B.~Sadoulet, and J.~Gao, ``Kinetic inductance phonon sensors for the
  cryogenic dark matter search experiment,'' \emph{J. Low Temp. Phys.}, vol.
  151, pp. 544--549, 2008.

\bibitem{golwala_jltp2008}
S.~Golwala, J.~Gao, D.~Moore, B.~Mazin, M.~Eckart, B.~Bumble, P.~Day, H.~LeDuc,
  and J.~Zmuidzinas, ``A wimp dark matter detector using mkids,'' \emph{J. Low
  Temp. Phys.}, vol. 151, pp. 550--556, 2008.

\bibitem{ekinci_roukes_nems_review}
K.~L. Ekinci and M.~L. Roukes, ``Nanoelectromechanical systems,'' \emph{Rev.
  Sci. Instrum.}, vol.~76, p. 061101, 2005.

\bibitem{schwab_roukes_phystoday2005}
K.~C. Schwab and M.~L. Roukes, ``Putting mechanics into quantum mechanics,''
  \emph{Phys. Today}, vol.~58, pp. 36--42, 2005.

\bibitem{kippenberg_vahala_science2008}
T.~J. Kippenberg and K.~J. Vahala, ``Cavity optomechanics: Back-action at the
  mesoscale,'' \emph{Science}, vol. 321, pp. 1172--1176, 2008.

\bibitem{marquardt_girvin_physics2009}
F.~Marquardt and S.~M. Girvin, ``Optomechanics,'' \emph{Physics}, vol.~2,
  p.~40, 2009.

\bibitem{grockowiak_sst2013}
A.~Grockowiak, T.~Klein, E.~Bustarret, J.~Ka{\v c}mar{\v c}{\' i}k, C.~Dubois,
  G.~Prudon, K.~Hoummada, D.~Mangelinck, T.~Kociniewski, D.~D{\' e}barre,
  J.~Boulmer, and C.~Marcenat, ``Superconducting properties of laser annealed
  implanted si:b epilayers,'' \emph{Supercond. Sci. Technol.}, vol.~26, p.
  045009, 2013.

\bibitem{muller}
R. Muller, private communication.

\end{thebibliography}







\end{document}